\documentclass[preprint,showpacs,showkeys,preprintnumbers,amsmath,amssymb,superscriptaddress]{revtex4}

\usepackage{graphicx}
\usepackage{dcolumn}
\usepackage{bm}
\DeclareMathOperator{\sech}{sech}

\begin{document}

\author{J. S. Oliveira Filho}
\affiliation{Departamento de F\'{\i}sica, Instituto de Ciências
Exatas, Universidade Federal de Minas Gerais, C.P. 702, 30161-970,
Belo Horizonte, MG, Brazil}
\author{R. Rossi Jr.}
\email{romeu.rossi@ufv.br}
\affiliation{Universidade Federal de Vi\c{c}osa, Campus Florestal,
C.P 35690-000 - Florestal, MG - Brasil}
\author{M. C. Nemes}
\affiliation{Departamento de F\'{\i}sica, Instituto de Ciências
Exatas, Universidade Federal de Minas Gerais, C.P. 702, 30161-970,
Belo Horizonte, MG, Brazil}

\title{Quantum Properties of a Which-Way Detector}

\begin{abstract}
We explore quantum properties of a which-way detector using three versions of an idealized two slit arrangements. Firstly we derive complementarity relations for the detector; secondly we show how the ''experiment'' may be altered in such a way that using single position measurement on the screen we can obtain quantum erasure. Finally we show how to construct a superposition of ``wave'' and ``particle'' components.
\end{abstract}

\maketitle

\section{Introduction}

Wave particle duality is one of the most counterintuitive features of quantum theory. In double slit experiments such problematic feature becomes explicit by the fact that if, in an experimental apparatus, the information about which slit the quanton (in the sense of \cite{art1,art2}) has crossed (which-way information) is available there is not interference fringes on the screen (particle behavior), however, if the which-way information is not available one can see an interference pattern (wave behavior). In a famous debate between N. Bohr and A. Einstein, at the Solvay conference (1927) \cite{art3}, they introduced a gedanken experiment which consists in a double slit experiment with a movable slit placed before the double slit, the Recoiling Slit Experiment. A quanton is first send through the movable slit, after it crosses the double slit and then it is recorded on the screen. Therefore, observations of the movable slit position, after the interaction with the quanton, can give us information about which slit (of the double slit apparatus) the quanton has crossed. The apparent difficulty imposed by such apparatus is that one could have both particle and wave behavior in the same experiment in contradiction to the wave particle duality. However this apparent difficulty was solved by N. Bohr which pointed out that a careful analysis of the movable slit would require the inclusion of uncertainty relations of its position and momentum, this would add random phases in the quantons path and consequentially it would make the interference pattern vanish. N. Bohr arguments in favor of the wave particle duality were based on the uncertainty principle.

Around the 80's entanglement starts to play an important role in the analysis of interferometric systems. In references \cite{art4, art5, art6} the authors show that the entanglement between an interferometric quanton and a meter (included in the quantum description) can destroy the interference pattern. The interaction with a probe system makes the which-way information available and it is sufficient to wash out the interference pattern. According to the authors, entanglement is determinant in this phenomena and it is not necessary to call upon Heisenberg's uncertainty principle as it was done in the early discussions between A. Einstein and N. Bohr. A debate on the role of entanglement and uncertainty relations begun \cite{art7, art8, art9, art10, art11}. It was also shown that the which-way information available in the entangled state can be erased, and consequently the interference pattern recovered. Experimental observations of the quantum eraser have been reported in several quantum systems \cite{art12, art13, art14, art15, art16, art17}.

The inclusion of the meter in quantum description of another famous experiment, the Wheeler's delayed choice experiment \cite{art18}, also provide new investigations on the wave-particle duality. Recently, several Quantum Delayed-Choice Experiment were proposed \cite{art19} and performed \cite{art20, art21, art22, art23}. Such experiments are based on the substitution of a ``classical beam splitter'' by a quantum system that can be prepared in a superposition state of being present or being absent. A  Mach-Zender interferometer is considered, therefore, the wave or particle behavior observed depends on the presence or absence of the second beam splitter. In Wheeler's delayed choice experiment a classical apparatus control the presence or absence of the second beam splitter after the insertion of the quanton in the apparatus. In the recent version, the Quantum Delayed-Choice Experiment, a quantum system plays the role of the second beam splitter, and it allows for the construction a ``superposition of wave and particle behavior''.

In the present work we explore the complementarity relations of measuring apparatus in the context os quantum erasure and quantum delayed choice. In the first section we study a two slit experiment already proposed in \cite{art24}. Now considering the complementarity relations of the detection system, two cavities, one before each slit, where the crossing atom leaves a photon which taggs the atom path. These two possibilities are then viewed as the two interferometric alternatives of the apparatus. In this context the atomic center of mass degree of freedom will play store which-way information of the cavities. Furthermore we show that a single measurement on the atomic position on the screen may generate a perfectly balanced superposition of the cavities state. Moreover we propose a gedanken experiment where the final cavities state may be interpreted as superposition of ``wave like'' and ``particle like'' states.

\section{Which-Way Detector}

\subsection{Non-selective Position Measurements}

Let us consider the double-slit experiment with high-Q cavities that work as which-way ``detectors'' proposed by M. O. Scully et al. \cite{art24}. Two level atoms prepared in the excited state cross, one at the time, a double-slit apparatus with a high-Q micromaser cavity placed on the entrance of each slit. The atom interacts with the cavity mode $M_{+}$ ($M_{-}$) before crossing the slit $+$ ($-$). The interaction time with the mode corresponds to a $\pi$ pulse, so that the atom leaves the excitation in the corresponding mode. The state vector of the global system after the double-slit is

\begin{equation}
|\psi\rangle = \left(\lambda_{+} |\psi_{+}\rangle|1_{+},0_{-}\rangle + \lambda_{-}e^{i\phi}|\psi_{-}\rangle|0_{+},1_{-}\rangle\right)|g\rangle,\label{ini}
\end{equation}
 where $\langle \psi_{+}|\psi_{+}\rangle=\langle \psi_{-}|\psi_{-}\rangle=1$, $\langle \psi_{+}|\psi_{-}\rangle=\langle \psi_{-}|\psi_{+}\rangle=e^{-\frac{d^{2}}{4b^{2}}}$, $\lambda_{+}^{2}+\lambda_{-}^{2}=1$, $|\psi_{+}\rangle$ ($|\psi_{-}\rangle$) are state vectors in the center-of-mass coordinate subsystem corresponding to the atom crossing slit $+$ ($-$), $d$ is the distance between the center of the slits and $b$ the initial wave packet width. If the atom crosses the double-slit through cavity $+$ ($-$) it leaves one excitation on mode $M_{+}$ ($M_{-}$), the which-way information of the atom is completely available in the cavity modes subsystem. Therefore, the modes $+$ ($-$) play the role of a which way detector. The quality of this detector is characterized by its complementarity relations as show in what follows.

In this analysis both the particle and the which way ``detector'' are included in the quantum description. The which-way information available in the detectors reduces the visibility of the interference pattern on the screen. The duality between which-way information and visibility is quantified by complementarity relations. In the present section we reverse the analysis and consider the interferometric properties of the ``detector'', i.e. we consider the cavity modes as a two way interferometer. In such interpretation the two interferometric alternatives are $|1_{+},0_{-}\rangle$ and $|0_{+},1_{-}\rangle$. The which-way information about these interferometric alternatives is available in the center-of-mass coordinate system of the two level atom.

To make the analysis concrete we consider the center-of-mass coordinate state of the quanton described by Gaussian wave packets

\begin{eqnarray}
|\psi_{+}\rangle=\int   \psi_{+}\left(x,t\right)\left|x\right\rangle dx &\\
|\psi_{-}\rangle=\int   \psi_{-}\left(x,t\right)\left|x\right\rangle dx,
\end{eqnarray}
where  $\psi_{\pm}\left(x,t\right)=\left[\frac{1}{B\left(t\right)\sqrt{\pi}}\right]^{\frac{1}{2}}exp\left[-\frac{\left(x\mp \frac{d}{2}-\frac{b^{2} k t}{\tau}\right)^{2}}{2 B^{2}\left(t\right)}\left(1-\frac{i t}{\tau}\right)-\frac{ik^{2}b^{2}t}{2 \tau} +ikx \right] $, $k$ is the transverse wave number, $d$ is the distance between the center of the slits, $b$ is the initial wave packet width, $B\left(t\right)=b\sqrt{1+\frac{t^{2}}{\tau^{2}}}$,  and the scale that characterize the variation of $B\left(t\right)$ is given by $\tau=\frac{m b^{2}}{\hbar}$, where $m$ is the quanton's mass. We consider a one dimensional wave packet which corresponds to the assumption that the spread of the wave packet is in the transverse direction to the beam propagation considered classical. This is justified provided the spread in this direction is sufficiently small, what can be achieved by a high enough longitudinal velocity.

If we ignore the atomic position measurements on the screen and trace over the continuous variable degree of freedom, we have the reduced state:

\begin{equation}
\rho_{S}=\left( \lambda_{+}^{2} |1_{+},0_{-}\rangle \langle 1_{+},0_{-}|+\lambda_{-}^{2} |0_{+},1_{-}\rangle \langle 0_{+},1_{-}|+\lambda_{+}\lambda_{-}e^{i \phi-\frac{d^{2}}{4 b^{2}}} |0_{+},1_{-}\rangle \langle 1_{+},0_{-}|+c.c.\right) |g\rangle \langle g|,
\end{equation}
where $Tr\left(\rho_{S}\right)=\lambda_{+}^{2}+\lambda_{-}^{2}=1$.

Using the definitions for visibility ($V$) and predictability ($P$) for a two level system \cite{art25} we have:

\begin{eqnarray}
P&=&\left|\lambda_{+}^{2}-\lambda_{-}^{2}\right|,\\
V&=&2 \left|\lambda_{+}\lambda_{-}\right| e^{-\frac{d^{2}}{4b^{2}}}.
\end{eqnarray}

A well known result is that the visibility and predictability are related by the inequality $V^{2}+P^{2} < 1$ when the interferometric system is not in a pure state. In Ref.\cite{art25} the authors show that the missing quantity that turns the inequality into an equality is the entanglement. Here we calculate the linear entropy, that quantifies the entanglement between the center of mass and cavity degrees of freedom, and obtain
\begin{equation}
S=2 \lambda_{+}^{2} \lambda_{-}^{2} \left(1 - e^{-\frac{d^{2}}{2b^{2}}}\right),
\end{equation}
and show that $P^{2}+V^{2}+2S=1$.

The linear entropy $S$ quantifies the which-way information available in the detector. Note that the vality of the model requires that the overlap between the two center of mass gaussian states is small enough, i. e., $e^{-d^{2}/(4b^{2})}\ll1$, yielding a neglible visibility. In this case it is ensured that the particle that goes through the slit $+$ ($-$) only interacted with mode $M_{+}$ ($M_{-}$).

We can define the distinguishability of the detector (associated with its quality) as \cite{art26}

\begin{equation}
D=\sqrt{P^{2}+2S}=\sqrt{1-V^{2}}\approx1.
\end{equation}

This can be illustrated by the two extreme situations. Let us assume that $S=0$. In this case we can be sure that the quanton crossed slit $+$ $(-)$, recorded by the detector final state $|1_{+},0_{-}\rangle$ $(|0_{+},1_{-}\rangle)$. When $P=0$, i. e., $S$ is maximum, the detector will be found in a statistical mixture which means that the quanton crossed the slits in a maximally coherent superposition of states $|\psi_{+}\rangle$ and $|\psi_{-}\rangle$ and are therefore completely entangled with the detector.

\subsection{Selective Position Measurement (Quantum Eraser)}

In this subsection we consider selective measurements of the center of mass position on a (distant) screen and show that some of these measurements correspond to a quantum erasure process for the detector subsystem in the sense that it will be left in a coherent superposition. Therefore it will necessarilly have a nonvanishing visibility. The common suport of the wave packets, that increases in time, decreases the ``quality'' of the which-way information encoded in the center of mass system. In the region where the two gaussian states are superimposed a measurement of the atomic position on the screen gives ambiguous information about the interferometric alternatives ($|1_{+},0_{-}\rangle$ and $|0_{+},1_{-}\rangle$). Therefore the such superposition allows us to perform measurements (on the observable $X$) which increase the visibility of the interferometric subsystem. These measurements can be interpreted as quantum eraser measurements, because they erase the which-way information and increase the visibility in the modes of the subsystem.

Let us consider that a measurement of the atomic position on the screen is performed and the eigenvalue $x$ is obtained. After the measurement the cavity modes state vector is given by:

\begin{equation}
|\psi(x,t)\rangle=\lambda_{+} \psi_{+}\left(x,t\right)|1_{+},0_{-}\rangle + \lambda_{-}e^{i\phi} \psi_{-}\left(x,t\right)|0_{+},1_{-}\rangle,
\end{equation}
with $\langle \psi_{+}|\psi_{+}\rangle=\langle \psi_{-}|\psi_{-}\rangle=1$.

To study the consequences of the atomic position measurements on the interferometric system let us consider the quantitative complementarity relation introduced in Ref.\cite{art25}

\begin{equation}
V^{2}_{x}+K^{2}_{x}\leq 1.\label{ine}
\end{equation}
To introduce such the inequality the authors consider a bipartite system composed by the interferometric system, with $|+\rangle$ and $|-\rangle$ as interferometric alternatives, and a which-way detector. $X$ is an observable in the which-way detector subsystem and $x$ is one eigenvalue of $X$. The inequality (\ref{ine}) is a quantitative complementarity relation for the interferometric system after the measurement of $X$ with the result $x$. $V_{x}$ is the ``conditioned visibility'' that depends on the choice of the measured observable ($X$) and on the obtained eigenvalue $x$. The ``conditioned visibility'' is calculated as the visibility  in the state vector after the measurement. The ``conditioned which-way knowledge'' $K_{x}$ reflects the a posteriori which-way knowledge (after the measurement) and is given by $K_{x}=|p(+|x)-p(-|x)|$, where $p(+|x)$ ($p(-|x)$) is the conditioned probability that the interferometric system took the alternative $|+\rangle$ ($|-\rangle$), conditioned that the eigenvalue $x$ has been obtained.

In the present system, the observable measured in the which-way detector subsystem is the atomic center-of-mass coordinated on the screen ($X$) with eigenvalues represented by $x$. We calculate the conditioned visibility ($V_{x}$) and knowledge ($K_{x}$)

\begin{eqnarray}
V_{x}=\frac{2\left|\lambda_{+}\right|\left|\lambda_{-}\right|}{\left|\lambda_{-}^{2}\left(\cosh\delta+\sinh\delta\right)+\lambda_{+}^{2}\left(\cosh\delta-\sinh\delta\right)\right|}\\
K_{x}=\frac{\left|\lambda_{-}^{2}\left(\cosh\delta+\sinh\delta\right)-\lambda_{+}^{2}\left(\cosh\delta-\sinh\delta\right)\right|}{\left|\lambda_{-}^{2}\left(\cosh\delta+\sinh\delta\right)+\lambda_{+}^{2}\left(\cosh\delta-\sinh\delta\right)\right|}
\end{eqnarray}
where $\delta=\frac{d\tau \left(b^{2}kt-x\tau\right)}{b^{2}(t^{2}+\tau^{2})}$.

The quantitative complementarity relation is then
\begin{equation}
V^{2}_{x}+K^{2}_{x}=1
\end{equation}
that corresponds to the inequality (\ref{ine}). The quantities $V_{x}$ and $K_{x}$ depends on the measured eigenvalue $x$ but it also depends on $t/\tau$. Let us consider that  $\lambda_{+}=\lambda_{-}=1/\sqrt{2}$. In this case, the expression for the conditioned visibility becomes

\begin{equation}
V_{x}= \sech\left\{\frac{d\tau \left(b^{2}kt-x\tau\right)}{b^{2}(t^{2}+\tau^{2})}\right\}
\label{vis}
\end{equation}

Equation (\ref{vis}) shows that the values of $x$ associated with the quantum eraser are given by $x=b^{2} k t/\tau$; notice that the complete quantum erasure occurs where the conditional visibility is maximum. For $k=0$ the complete quantum erasure occurs for the measurement on the screen $x=0$, independently of $b$ and time $t/\tau$. However, if $k\neq0$ and $t/\tau\neq0$, the system is no more symmetric and the values of $x$ for a complete quantum erasure depend on the time.

Firstly let us analyze the position $x$ dependence. In Fig.1 it is shown a curve of the conditioned visibility ($V_{x}$) and the knowledge ($K_{x}$) as a function of $x/b$ for a fixed time of propagation $t/\tau=1$. We can see that if the atom is measured away form the center of the screen $V_{x}$ is approximately zero and the $K_{x}$ is approximately maximum. In these regions of the screen there is no superposition between the Gaussian wave packets, so there is a high probability to make a right guess about which slit the atom has crossed. On the other hand, if the atom is measured in the center of the screen, the interferometric system is projected onto a state with maximum conditioned visibility, therefore such measurement work as a quantum eraser. It washes out the conditioned knowledge about the interferometric alternatives. It is interesting to notice that one can observe the continuous variation of $V_{x}$, from the maximum to the minimum value, just with measurements of one observable $X$. Measurements of the same observable $X$ either provides information of which path the atom crossed $(\left|x\right|/b\gg0)$ or works as quantum erasure $(\left|x\right|/b\approx0)$ \cite{art25}

\begin{figure}[h]
\centering
  \includegraphics[scale=0.6]{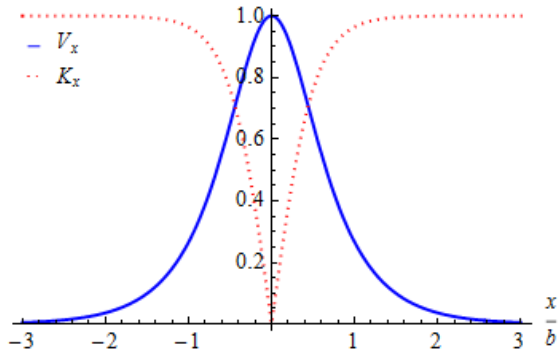}\\
\caption{Conditional visibility as a function of the position on the screen, with $\lambda_{+}=\lambda_{-}=1/\sqrt{2}$, $k =0$, $d/b=4$, $t/\tau=1$.}
\end{figure}\label{fig1}

In Fig.2 we show the conditioned visibility ($V_{x}$) and the knowledge ($K_{x}$) as a function of the propagation time $t/\tau$ for a fixed position on the screen $x/b=1$. Notice that for small propagation time the conditional visibility is close to zero and it increases over time. Therefore, for small propagation time (up to $t/\tau\approx1$) if the quanton is detected on the screen at position $x/b=1$, the probability that it has crossed slit $+$ is very high. However, over time the which-way information of the quanton measured in $x/b=1$ becomes ambiguous and the conditional visibility increases. In the present system the same measurement result $x$ works initially as a which-way sorting (up to $t/\tau\approx1$) and later as a quantum eraser sorting.

\begin{figure}[h]
\centering
  \includegraphics[scale=0.6]{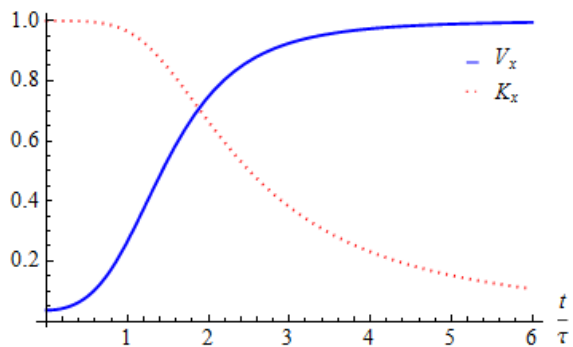}\\
\caption{Conditional visibility as a function of the time, with $\lambda_{+}=\lambda_{-}=1/\sqrt{2}$, $k=0$, $d/b=4$, $x/b=1$.}
\end{figure}\label{fig2}

In Fig.3 we use the three dimensional plot when it is clearly seen that there exists a straight line $x=b^{2}kt/\tau$ where the quantum conditioned visibility is maximum for fixed $t/\tau=1$.

\begin{figure}[h]
\centering
  \includegraphics[scale=0.6]{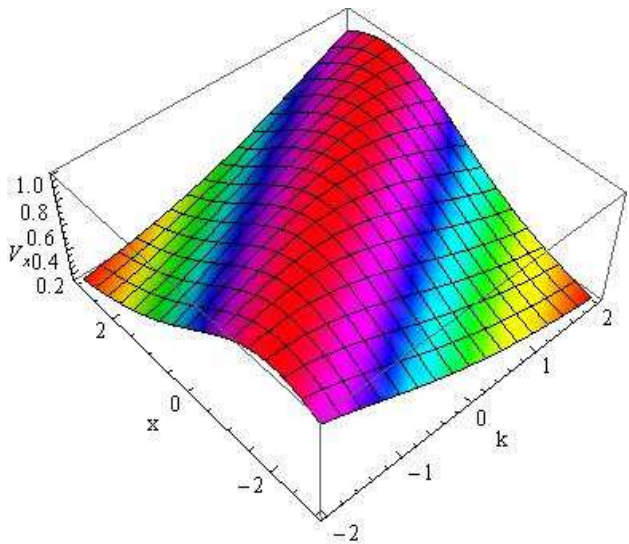}\\
\caption{Conditional visibility as a function of position $(x)$ and the transverse wave number ($k$), with  $\lambda_{+}=\lambda_{-}=1/\sqrt{2}$, $d/b=4$ and $t/\tau=1$.}
\end{figure}\label{fig3}

\section{Wave and Particle Superposition}

We now turn our attention to the construction an entangled state. The coefficient which is composed by another degree of freedom of one of them is a superposition state (``wave''). On the other hand, the ``particle behavior'' is associated with a product state. We propose the construction of such state in the cavity system, but we slightly change the apparatus described in the last section. We consider now a double-slit experiment with only one high-Q cavity, that is placed on the entrance of slit $+$. We also include a Ramsey Zone that can rotate the internal atomic state next to the high-Q cavity. Therefore, the two level atoms that cross the slit $+$ interact with mode $M_{+}$ inside the high-Q cavity and after interaction with a classical electromagnetic field inside the Ramsey Zone. The atom that crosses the slit $-$ does not interact with any electromagnetic field.

\begin{figure}[h]
\centering
  \includegraphics[scale=0.6]{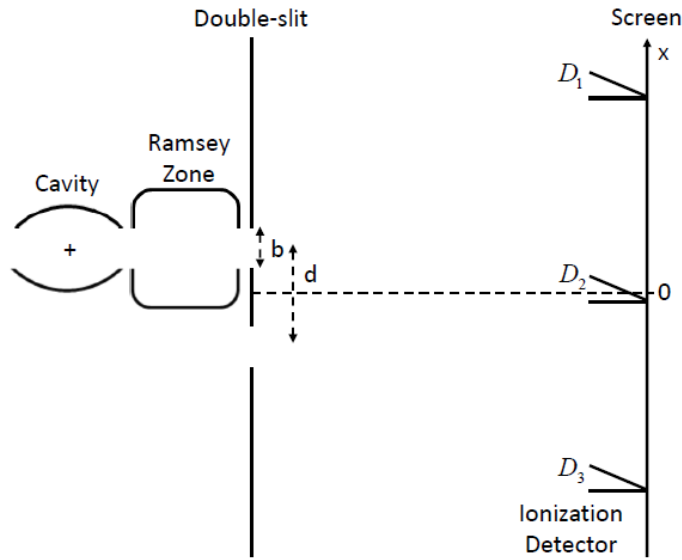}\\
\caption{Schematic sketch of the proposed apparatus.}
\end{figure}\label{fig4}

The mode $M_{+}$ is an interferometric system with the two ortogonal ``paths'' $|0_{+}\rangle$ and $|1_{+}\rangle$. When a state vector of $M_{+}$ is in a maximum superposition state of the alternatives $|0_{+}\rangle$ and $|1_{+}\rangle$, we can consider that it exhibits wave properties. On the other hand, if $M_{+}$ is in a well defined ``path" state $|0_{+}\rangle$ or $|1_{+}\rangle$ it exhibits particle properties. In the present section we present a scheme to prepare a global state where each Gaussian wave packet is associated to a specific character (wave or particle) of $M_{+}$. More specifically, the scheme associates $|\psi_{+}\rangle$ to the wave character and $|\psi_{-}\rangle$ to the particle character.

We consider that the interaction time between the atom and $M_{+}$ corresponds to a $\pi/2$ pulse. The atom that crosses a slit $+$ interacts first with mode $M_{+}$ and evolves as
\begin{equation}
|\psi_{+}\rangle|0_{+}\rangle|e_{+}\rangle\rightarrow|\psi_{+}\rangle\frac{1}{\sqrt{2}}\left(|0_{+}\rangle|e_{+}\rangle+|1_{+}\rangle|g_{+}\rangle\right).
\end{equation}
 In the Ramsey Zone the atom in ground state evolves as $|g\rangle\rightarrow|g\rangle-|e\rangle$ and the excited state as $|e\rangle\rightarrow|g\rangle+|e\rangle$.

The state vector of the global system after the double-slit is given by
\begin{equation}
|\psi\rangle=|\psi_{+}\rangle\left[\frac{1}{\sqrt{2}}\left(|0_{+}\rangle|e_{+}\rangle+|0_{+}\rangle|g_{+}\rangle\right)+ \frac{1}{\sqrt{2}}\left(|1_{+}\rangle|g_{+}\rangle-|1_{+}\rangle|e_{+}\rangle\right)\right]+|\psi_{-}\rangle|0_{+}\rangle|e_{+}\rangle.\label{sta2}
\end{equation}

Suppose that it is possible to measure the atomic energy and the atomic position on the screen. To simplify the imaginary apparatus let us consider three atomic ionization detectors, as it is shown in Fig.4. If we consider only the detections of the excited state $|e\rangle$, we obtain only in the following part of the state \ref{sta2}

\begin{equation}
|\psi_{+}\rangle\frac{1}{\sqrt{2}}\left(|0_{+}\rangle -|1_{+}\rangle\right)+|\psi_{-}\rangle|0_{+}\rangle.\label{sta3}
\end{equation}

The Gaussian wave packet $|\psi_{+}\rangle$ is associated with a maximum visibility state (``wave'' state) and the Gaussian wave packet $|\psi_{-}\rangle$ is associated with a maximum predicability state (``particle'' state). Consider that the screen is positioned at a fixed distance. Therefore, the clicks on the detectors shown in Fig.4 will be responsible for the preparation of subsystem $M_{+}$ on a ``wave'' state, ``particle'' state or superposition of ``wave'' and ``particle'' state. If $D_{1}$ clicks the state on the mode $M_{+}$ can be written as $\frac{1}{\sqrt{2}}\left(|0_{+}\rangle -|1_{+}\rangle\right)=|w\rangle$, because in this region of the space there is no common support between $|\psi_{+}\rangle$ and $|\psi_{-}\rangle$, and we can consider that the detected atom had crossed slit $+$. If $D_{1}$ or $D_{3}$ clicks the state of mode $M_{+}$ can now be written as $|0_{+}\rangle=|p\rangle$, because the detected atom had crossed slit $-$. However, if the atomic ionization detector is placed in a region of the space with a significant common support, we can not guarantee that the detected atom had crossed slit $+$ or $-$. When detector $D_{2}$ (which is in the center of the screen) clicks, the state on the mode $M_{+}$ can be written as $\frac{1}{\sqrt{2}}\left(|0_{+}\rangle -|1_{+}\rangle\right)+|0_{+}\rangle=|w\rangle+|p\rangle$ that corresponds to a superposition of ``wave'' and ``particle'' state.

In conclusion we have explore the quantum interferometric properties of subsystems that work as auxiliary (two cavity modes) in quantum eraser scheme. We wrote the quantitative complementarity relation for the two cavity modes subsystem, and also show that a single position measurement of the quanton on the screen can work as a quantum eraser and restore the visibility on the two modes cavity subsystem. We emphasize that the same measurement result of the quanton's position on the screen can work as a quantum eraser measurement or a which-way measurement, depending on the relations between the position measured and the propagation time of the particle. Finally we show a scheme to construct a ``wave''and ``particle'' superposition in a cavity mode subsystem using the measurement of the position on the screen of a two level atom.

\end{document}